\documentclass[%
 reprint,%
 amssymb, amsmath,%
 pra,cha,%
]{revtex4-1}
\usepackage{graphics}
\usepackage{amsmath}
\usepackage{amsfonts}
\usepackage{latexsym}
\usepackage{amsbsy}
\usepackage{epstopdf}
\usepackage{bm}%

\newcommand{\ket}[1]{\vert #1 \rangle}

\usepackage{color}

\usepackage{ulem} %for strikeout "\sout"

\newcommand{\dyadic}[1]{{#1}
\setbox0=\hbox{$\scriptstyle\leftrightarrow$}
   \setbox2=\hbox{$#1$}
   \dimen0=.5\wd0 \advance\dimen0 by-.5\wd2
   \advance\dimen0 by-\wd0
   \kern\dimen0
{^{\hbox{$\scriptstyle\leftrightarrow$}}}}

\begin{document}

%\title{Sub-Wavelength Imaging and Field Mapping via EIT and Autler-Townes Splitting In Rydberg Atoms}
%
%\author{~Christopher~L.~Holloway,~\IEEEmembership{Fellow,~IEEE,}
%        Josh A. Gordon, Andrew Schwarzkopf, David A. Anderson, \\ Stephanie A. Miller, Nithiwadee Thaicharoen, Georg Raithel
%        \thanks{Manuscript received \today.}\thanks{C.L. Holloway and J. Gordon are with the National
%Institute of Standards and Technology (NIST), Electromagnetics Division,
%U.S. Department of Commerce, Boulder Laboratories,
%Boulder,~CO~80305. A. Schwarzkopf, D. A. Anderson, S. A. Miller, N. Thaicharoen, and G. Raithel are with the Department of Physics, University of Michigan, %Ann Arbor, MI 48109.  This work was partially supported by DARPA's QuASAR program. Publication of the U.S. government, not subject to U.S. copyright.}}%
%
%\maketitle

%Independent RF Electric Field Metrology via Electromagnetically Induced Transparency Using Cs and Rb Rydberg Atoms Simultaneously\\

\title{Simultaneous Use of Cs and Rb Rydberg Atoms for Independent RF Electric Field Measurements via Electromagnetically Induced Transparency}
\thanks{Publication of the U.S. government, not subject to U.S. copyright. This work was partially supported by the Defense Advanced Research Projects Agency (DARPA) under the QuASAR Program and by NIST through the Embedded Standards program.}
\author{Matt T. Simons}
\author{Joshua A. Gordon}
\author{Christopher~L.~Holloway}
\email{holloway@boulder.nist.gov}
\affiliation{National Institute of Standards and Technology (NIST), Electromagnetics Division,
U.S. Department of Commerce, Boulder Laboratories,
Boulder,~CO~80305}

\date{\today}

\begin{abstract}
We demonstrate simultaneous electromagnetically-induced transparency (EIT) with cesium (Cs) and rubidium (Rb) Rydberg atoms in the same vapor cell with coincident (overlapping) optical fields. Each atomic system can detect radio frequency (RF) electric (E) field strengths through modification of the EIT signal (Autler-Townes (AT) splitting), which leads to a direct SI traceable RF E-field measurement. We show that these two systems can detect the same the RF E-field strength simultaneously, which provides a direct in situ comparison of Rb and Cs RF measurements in Rydberg atoms.  In effect, this allows us to perform two independent measurements of the same quantity in the same laboratory, providing two different immediate and independent measurements. This gives two measurements that helps rule out systematic effects and uncertainties in this E-field metrology approach, which are important when establishing an international measurement standard for an E-field strength and is a necessary step for this method to be accepted as a standard calibration technique. We use this approach to measure E-fields at 9.2~GHz, 11.6~GHz, and 13.4~GHz, which correspond to three different atomic states (different principal atomic numbers and angular momentums) for the two atom species.
\vspace{7mm}
\end{abstract}

%\pacs{}
%
%%%\keywords{atom based metrology, Autler-Townes Splitting, broadband probe, electrical field measurements and sensors, EIT, sub-wavelength imaging, Rydberg %%%atoms}

\maketitle

\section{Introduction}

A stated goal of international metrology organizations and National Metrology Institutes (NMIs), including the National Institute of Standards and Technology (NIST), is to make all measurements directly traceable to the International System of Units (SI). Whenever possible, we would like these metrology techniques to be able to make an absolute measurement of a physical quantity of interest plus any measurement based on the atom to provide a direct SI traceability path and hence enable absolute measurements of physical quantities. Measurement standards based on atoms have been used for a number of years for a wide array of applications; most notable are time, frequency, and length.  There is a need to extend these atom-based techniques to other physical quantities, such as electric (E) fields.  In recent work, we (and others) have demonstrated a fundamentally new approach for self-calibrated SI-traceable E-field measurements with the capability of fine spatial resolution (including sub-wavelength resolution) \cite{r1}-\cite{r5}.

\begin{figure}[!t]
\centering
%\scalebox{.33}{\includegraphics*{Vapor-Cell-both-levels_eps.eps}}
\scalebox{.35}{\includegraphics*{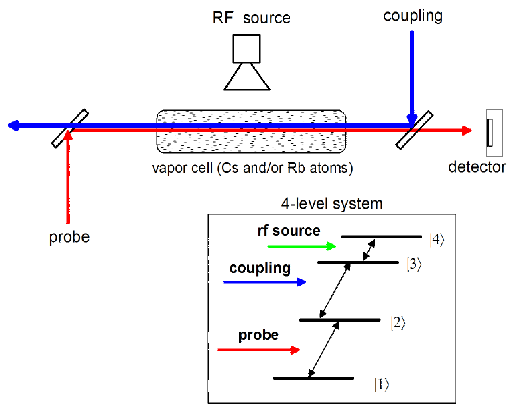}}
\caption{Illustration of a four-level system, and the vapor cell setup for measuring EIT, with counter-propagating probe and coupling beams. The RF is applied transverse to the optical beam propagation in the vapor cell.}
\label{4level}
\end{figure}

This new approach utilizes the concept of electromagnetically induced transparency (EIT) \cite{r1,r2, EIT_Adams}. Consider a sample of stationary four-level atoms illuminated by a single weak (``probe") light field, as depicted in Fig~\ref{4level}. In this approach, one laser is used to probe the response of the atoms and a second laser is used to excite the atoms to a Rydberg state (the “coupling” laser). In the presence of the coupling laser, the atoms become transparent to the probe laser transmission (this is the concept of EIT). The coupling laser wavelength is chosen such that the atom is at a high enough principle-quantum state such that an RF field can cause an atomic transition. The RF transition in this four-level atomic system causes Autler-Townes (AT) splitting of the transmission spectrum (the EIT signal) for a probe laser.  This splitting of the probe laser spectrum is easily measured and is directly proportional to the applied RF E-field amplitude (through Planck's constant and the dipole moment of the atom). By measuring this splitting, we can directly measure the RF E-field strength with the following \cite{r1}:
\begin{equation}
	|E| = 2 \pi \frac{\hbar}{\wp} \frac{\lambda_p}{\lambda_c} \Delta f_m= 2 \pi \frac{\hbar}{\wp}\Delta f_o \quad ,
	\label{mage2}
\end{equation}
where $\Delta f_m$ is the measured splitting and $\Delta f_o=\frac{\lambda_p}{\lambda_c} \Delta f_m$, $\hbar$ is Planck's constant and $\wp$ is the atomic dipole moment of the RF transition. The ratio $\frac{\lambda_p}{\lambda_c}$ (where $\lambda_p$ and $\lambda_c$ are the wavelengths of the probe and coupling lasers, respectively)  accounts for the Doppler mismatch of the probe and coupling lasers \cite{EIT_Adams}. We consider this type of measurement of the E-field strength a direct SI-traceable self-calibrated measurement in that it is related to Planck's constant (which will become an SI-defined quantity by standard bodies in the near future) and only requires a frequency measurement ($\Delta f_m$, which can be measured very accurately). A typical measured EIT signal from this technique is shown in Fig.~\ref{eitfig1} for the case with and without RF applied. The experimental setup and details are given below.  Application of RF (via a horn antenna placed 318~mm from the vapor cell) at 13.404~GHz couples two high laying Rydberg states and splits the EIT peak as shown in the solid curve in the figure.  We measured the AT splitting ($\Delta f_m$) of the EIT signal in the probe spectrum for a range of RF source levels, and determine the E-field amplitude using (\ref{mage2}). These values are also shown in the figure.

\begin{figure}
\centering
%\scalebox{.28}{\includegraphics*{EIT-singal-Cs-66S-66P_eps.eps}}\\
%{\tiny{(a)}}\\
%\scalebox{.28}{\includegraphics*{Power-Scans-0f-CS-43D-44P_eps.eps}}\\
%{\tiny{(b)}}
\scalebox{.31}{\includegraphics*{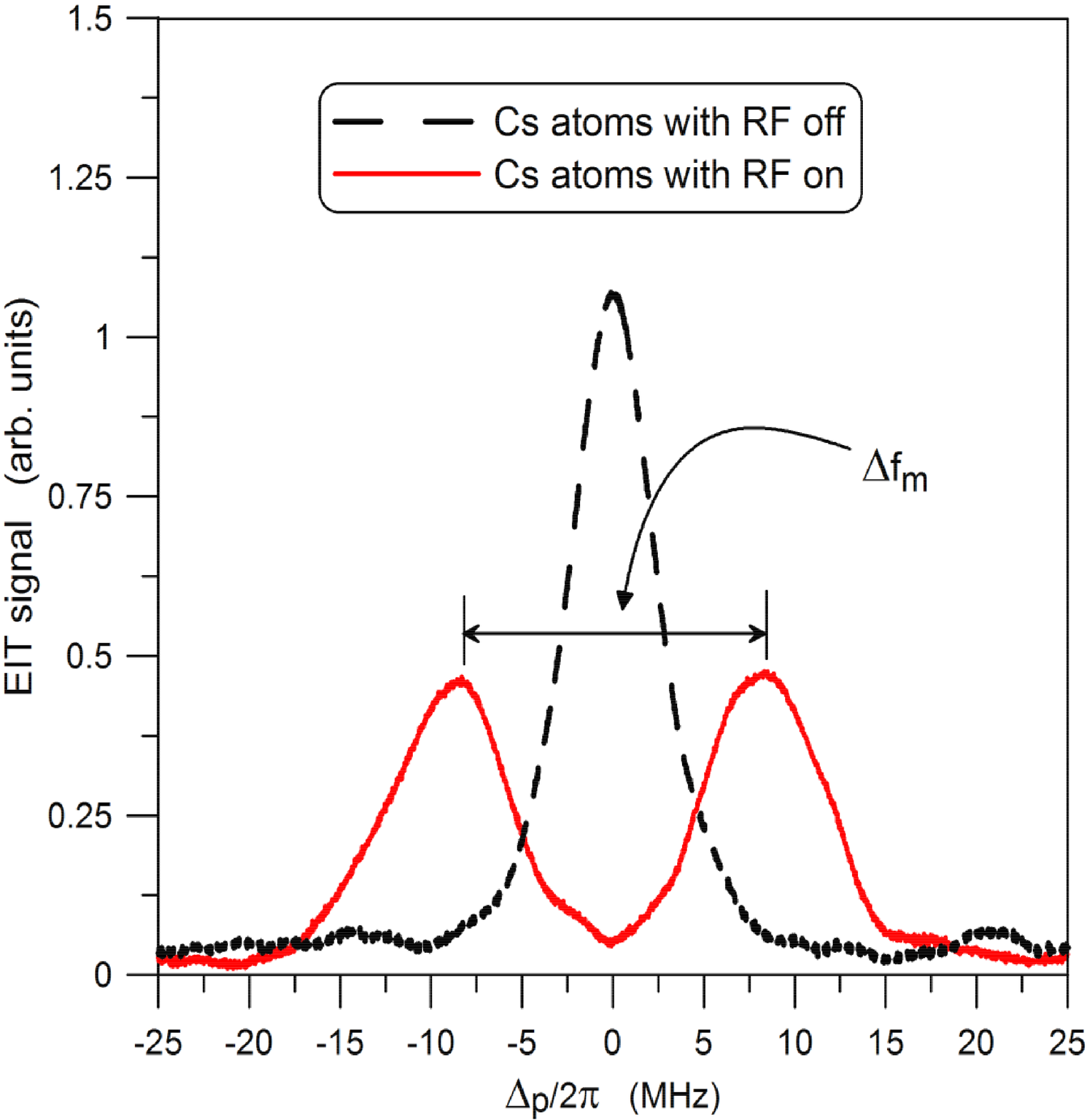}}\\
{\tiny{(a)}}\\
\scalebox{.31}{\includegraphics*{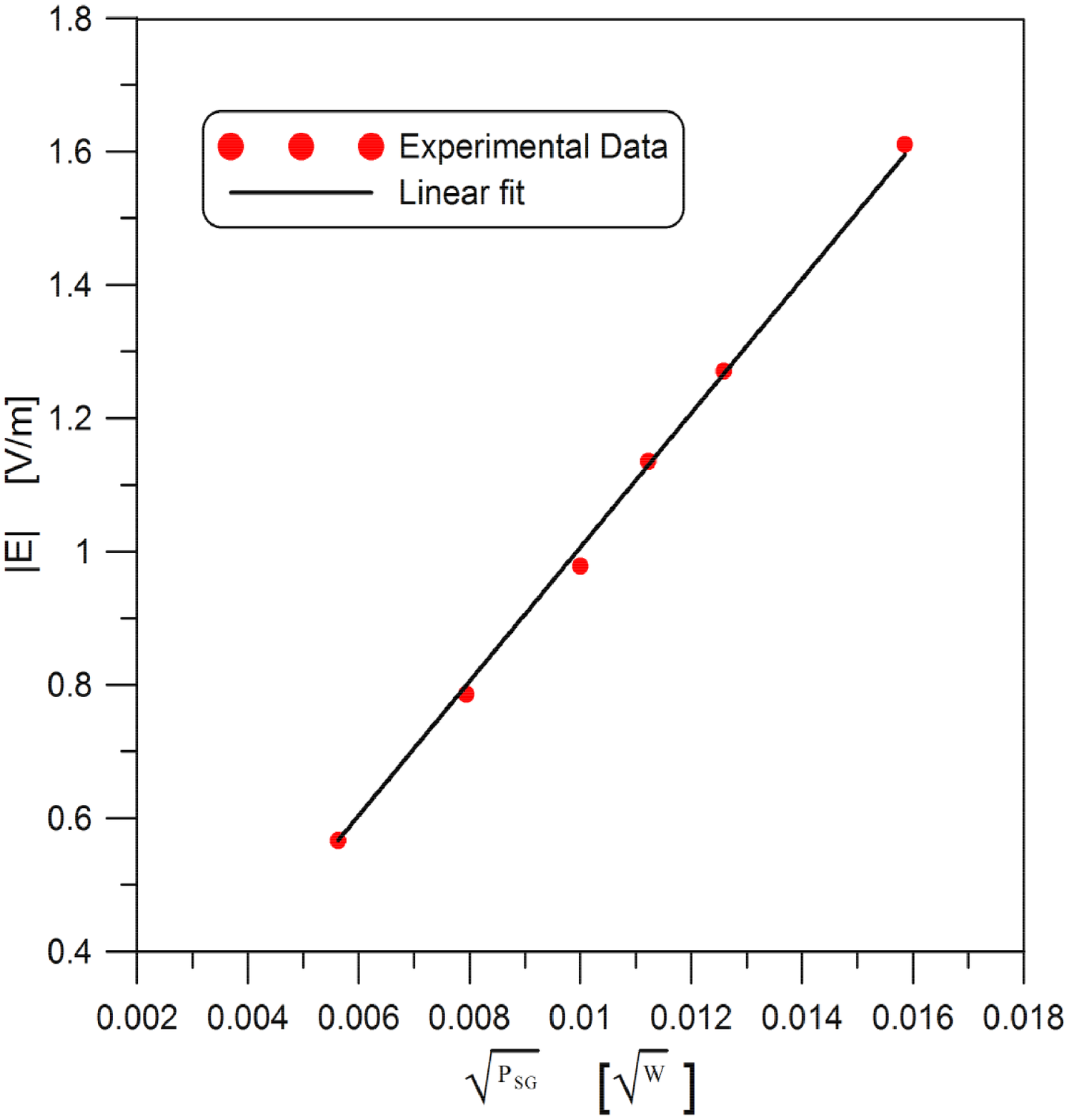}}\\
{\tiny{(b)}}
\caption{This dataset is for an RF of 13.404~GHz and corresponds to the following 4-level Cs atomic system: $6S_{1/2}-6P_{3/2}-66D_{5/2}-66P_{3/2}$. (a) Illustration of the EIT signal (i.e., probe laser transmission through the cell) as a function of probe laser detuning $\Delta p$.   (b) Calculated E-field for a given signal generator power $P_{SG}$, see discussion below.}
\label{eitfig1}
\end{figure}

%\begin{figure}
%\centering
%\scalebox{.3}{\includegraphics*{EIT-RF-on-off_eps.eps}}
%\caption{Estimated E-field for a 132.6459~GHz and corresponds to this following 4-level atomic system: $5S_{1/2}-5P_{3/2}-26D_{5/2}-27P_{3/2}$.}
%\label{powerscam}
%\end{figure}

The uncertainties of these types of measurements are currently being investigated \cite{r1, emc, fan}. With that said, for a new measurement method to be accepted by NMIs, the accuracy of the approach must be assessed. By performing simultaneous EIT measurements with two different atomic species in the same vapor cell with coincident (overlapping) optical fields exposed to the identical E-field, we can assess various aspects of the technique. In effect, this allows us to perform the same measurement in two different laboratories simultaneously, providing two independent measurements of the same E-field. There are subtle aspects of this technique that using two different atoms allows us to address and performing such dual atom experiments help in understanding systematic effects and uncertainties of this approach. For example, these experiments will help in assessing the accuracy of the dipole moment calculations of the various atoms.  In this paper, we demonstrate simultaneous E-field measurement via EIT using both cesium  atoms ($^{133}$Cs) and rubidium-85 ($^{85}$Rb) atoms in the same vapor cell. We discuss various aspects of these coincident tests by measuring E-fields in the 9.2~GHz, 11.6~GHz,  and 13.4~GHz frequency range.

\section{RF Transitions for Cs and Rb}

The broadband nature of this technique is due to the large number of possible Rydberg states that can exhibit a large response to an RF source \cite{r1}. In order to perform these types of simultaneous measurements, we need to choose states for $^{133}$Cs and $^{85}$Rb that have similar RF transition frequencies. While there are a large number of possible atomic states with RF transition frequencies, several of these have small atomic dipole moments. Since the measurement splitting ($\Delta f_m$) is directly proportional to the atomic dipole moments, we need to use RF transitions with large dipole moments.  The four classes of RF transitions corresponding to $nD_{5/2}$-$(n+1)P_{3/2}$, $nD_{5/2}$-$(n-1)F_{7/2}$, $nS_{1/2}$-$nP_{3/2}$, and $nS_{1/2}$-$(n-1)P_{3/2}$ have the largest dipole moments and are good choices for these experiments.  Table \ref{t1} shows a few of the possible states for $^{133}$Cs and $^{85}$Rb that exhibit similar RF transitions. The on-resonant RF transition frequencies are denoted as $f_{RF,o}$, which were obtained with the Rydberg formula and the quantum defects for $^{85}$Rb and $^{133}$Cs \cite{gal}-\cite{qdcsl}.  Also, in this table are the dipole-moments for each state, composed of a radial part ${\cal R}$ and an angular part ${\cal A}$, where $\wp={\cal R}{\cal {A}}$. The radial part ${\cal R}$ is obtained from two numerical calculations (see \cite{r1}) using the quantum defects for $^{85}$Rb and $^{133}$Cs \cite{gal}-\cite{qdcsl}. The angular part of the dipole moment is independent of $n$; for these four transitions (for $m_j=\pm1/2$), ${\cal A}_{34}=0.4899$ (for $nD_{5/2}$-$(n+1)P_{3/2}$), ${\cal A}_{34}=0.4949$ (for $nD_{5/2}$-$(n-1)F_{7/2}$), ${\cal A}_{34}=0.4714$ (for $nS_{1/2}$-$nP_{3/2}$), and ${\cal A}_{34}=0.4714$ (for $nS_{1/2}$-$(n-1)P_{3/2}$), see \cite{sobelman}. Note that these ${\cal A}$ correspond to co-linear polarized optical beams and the RF source, which is the case used in these experiments. In this table, we also show the percentage difference in the transition frequencies ($\%f$) between the Cs and Rb states, indicating that these six states have relatively close transition frequencies.

\begingroup
\begin{table}
\caption{RF Transitions for $^{133}$Cs and $^{85}$Rb. In the table, $e$ is the elementary charge and $a_o$ is the Bohr radius.}
\tiny
\label{t1}
\begin{center}
\begin{tabular}{|c||c|c|c|}\hline
& $^{133}$Cs states & $^{85}$Rb states & $\% f$\\
  \hline
1 & $47D_{5/2}-48P_{3/2}$  & $69D_{5/2}-68F_{7/2}$&  \\
 & $f_{RF,o}$=6.9458 GHz  & $f_{RF,o}$=6.9571 GHz  & 0.09~$\%$ \\
  & ${\cal R}_{Cs}$=2946.282\,${ea_o}$  & ${\cal R}_{Rb}$=6134.212\,${ea_o}$  &  \\
  & ${\cal A}_{Cs}$=0.4899  & ${\cal A}_{Rb}$=0.4949  & \\ \hline

2 & $45D_{5/2}-46P_{3/2}$  & $66D_{5/2}-65F_{7/2}$&  \\
 & $f_{RF,o}$=7.9752 GHz  & $f_{RF,o}$=7.96823 GHz  & 0.16~$\%$ \\
  & ${\cal R}_{Cs}$=2687.518\,${ea_o}$  & ${\cal R}_{Rb}$=5606.661\,${ea_o}$  &  \\
  & ${\cal A}_{Cs}$=0.4899  & ${\cal A}_{Rb}$=0.4949  &  \\ \hline

3 & $43D_{5/2}-44P_{3/2}$  & $61D_{5/2}-62P_{3/2}$&  \\
 & $f_{RF,o}$=9.2186 GHz  & $f_{RF,o}$=9.2264 GHz  & 0.01~$\%$ \\
  & ${\cal R}_{Cs}$=2440.629\,${ea_o}$  & ${\cal R}_{Rb}$=4829.407\,${ea_o}$  &  \\
  & ${\cal A}_{Cs}$=0.4899  & ${\cal A}_{Rb}$=0.4899  &  \\ \hline

4 & $40D_{5/2}-41P_{3/2}$  & $68S_{1/2}-68P_{3/2}$&  \\
 & $f_{RF,o}$=11.6187 GHz  & $f_{RF,o}$=11.6665 GHz  & 0.33~$\%$ \\
  & ${\cal R}_{Cs}$=2092.565\,${ea_o}$  & ${\cal R}_{Rb}$=4781.494\,${ea_o}$  &  \\
  & ${\cal A}_{Cs}$=0.4899  & ${\cal A}_{Rb}$=0.4714  &  \\ \hline

5 & $66S_{1/2}-66P_{3/2}$  & $65S_{1/2}-65P_{3/2}$&  \\
 & $f_{RF,o}$=13.4078 GHz  & $f_{RF,o}$=13.4398 GHz  & 0.20~$\%$ \\
  & ${\cal R}_{Cs}$=4360.132\,${ea_o}$  & ${\cal R}_{Rb}$=4352.837\,${ea_o}$  &  \\
  & ${\cal A}_{Cs}$=0.4714  & ${\cal A}_{Rb}$=0.4714  &  \\ \hline

6 & $63S_{1/2}-63P_{3/2}$  & $53D_{5/2}-52F_{7/2}$&  \\
 & $f_{RF,o}$=15.5513 GHz  & $f_{RF,o}$=15.5924 GHz  & 0.26~$\%$ \\
  & ${\cal R}_{Cs}$=3951.355\,${ea_o}$  & ${\cal R}_{Rb}$=3593.807\,${ea_o}$  &  \\
  & ${\cal A}_{Cs}$=0.4714  & ${\cal A}_{Rb}$=0.4949  &  \\ \hline

\end{tabular}
\end{center}
\end{table}
\endgroup

\section{Dual Atom Experimental Setup}

\begin{figure}
\centering
%\scalebox{.35}{\includegraphics*{Vapor-Cell_Cs-Rb_eps.eps}}
\scalebox{.35}{\includegraphics*{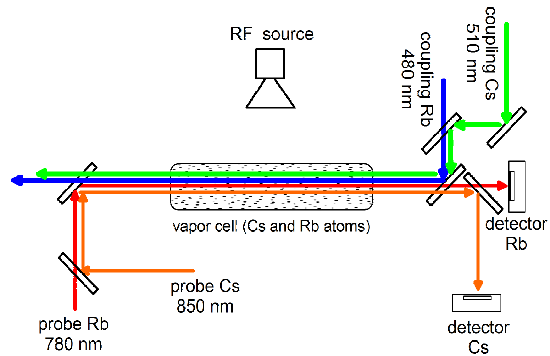}}
\caption{Illustration of the vapor cell setup for two atomic species EIT experiments, with counter-propagating probe and coupling beams. The RF is applied transverse to the optical beam propagation in the vapor cell. There are two probe beams (corresponding to Cs and Rb) and two coupling beams (for Cs and Rb) with all four beams overlapping. {In the diagram the beams are separated for clarity; in reality they are all coincident.}}
\label{CsRb-setup}
\end{figure}

The experimental setup is shown in Fig.~{\ref{CsRb-setup}}. We use a cylindrical glass vapor cell of length 75~mm and diameter 25~mm containing both $^{85}$Rb atoms and $^{133}$Cs.  For the $^{85}$Rb atoms, the levels $\ket{1}$, $\ket{2}$, $\ket{3}$, and $\ket{4}$ correspond respectively to the $^{85}$Rb  $5S_{1/2}$ ground state,  $5P_{3/2}$ excited state, and two Rydberg states.  The probe for $^{85}$Rb is a 780.24~nm laser which is scanned across the $5S_{1/2}$ -- $5P_{3/2}$ transition and is focused to a full-width at half maximum (FWHM) of 80~$\mu$m, with a power of 120~nW. To produce an EIT signal in $^{85}$Rb, we apply a counter-propagating coupling laser (wavelength $\lambda_c \approx 480$~nm) with a power of 32~mW, focused to a FWHM of 144~$\mu$m.   For the $^{133}$Cs atoms, the levels $\ket{1}$, $\ket{2}$, $\ket{3}$, and $\ket{4}$ correspond respectively to the $^{133}$Cs  $6S_{1/2}$ ground state,  $6P_{3/2}$ excited state, and two Rydberg states.  The probe for $^{133}$Cs is a 850.53~nm laser which is scanned across the $6S_{1/2}$ -- $6P_{3/2}$ transition and is focused to a full-width at half maximum (FWHM) of 80~$\mu$m, with a power of 120~nW. To produce an EIT signal in $^{133}$Cs, we apply a counter-propagating coupling laser (wavelength $\lambda_c \approx 510$~nm) with a power of 32~mW, focused to a FWHM of 144~$\mu$m.  {In order to ensure both Cs and Rb see the same RF field, all four beams are overlapped and focused on to the same spot inside the vapor cell using a bean profiler.}  We used two different photodetectors (one for the Rb atoms and one for the Cs atoms), allowing us to measure the EIT signal for both atoms separately and/or simultaneously.  We modulate the coupling lasers' amplitude with a 30~kHz square wave and detect any resulting modulation of the probe transmission with a lock-in amplifier. This removes the Doppler background and isolates the EIT signal. The RF E-field at the vapor cell was applied by a signal generator (SG) connected to a horn antenna via an RF cable.  The RF power levels ($P_{SG}$) stated in this paper are the power readings of the SG that feeds the cable which, in turn, feeds the horn antenna.  Due to the losses in the feeding cable, the reflections and losses in the horn antenna, and the propagation losses, this is not the power levels (or $E$-field strengths) incident onto the vapor cell. The E-field strength at the vapor cell is determined by taking into account these various losses.

\subsection{$^{133}$Cs: $43D_{5/2}$-$44P_{3/2}$  and  $^{85}$Rb: $61D_{5/2}$-$62P_{3/2}$}

\begin{figure}[!h]
\centering
%\scalebox{.28}{\includegraphics*{EIT-singal-CsandRb-p6dBm-TWO_eps.eps}}
\scalebox{.33}{\includegraphics*{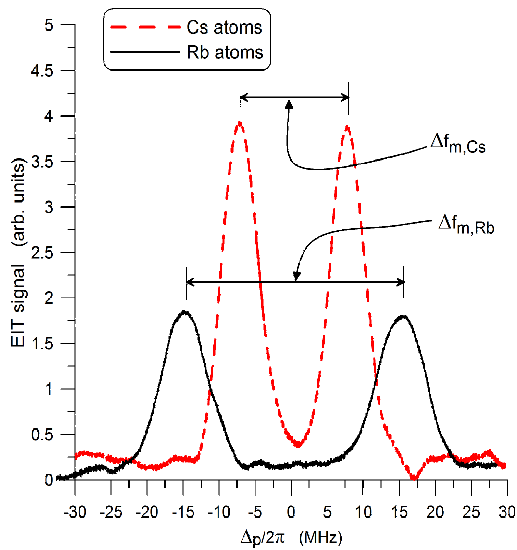}}
\caption{Illustration of the EIT signal (i.e., probe laser transmissions through the cell) as a function of probe laser detuning $\Delta_p$.  The dashed line is the measured EIT signal obtained with the CS atoms and the solid line is the measured EIT signal obtained with the Rb atoms.  The observed splitting result from an applied 9.222 GHz RF E-field.}
\label{CsRbEIT}
\end{figure}

We first performed experiments for an RF transition of approximately 9.22~GHz. From Table \ref{t1}, we see that this transition corresponds to $6S_{1/2}$-$6P_{3/2}$-$43D_{5/2}$-$44P_{3/2}$ for $^{133}$Cs and $5S_{1/2}$-$5P_{3/2}$-$61D_{5/2}$-$62P_{3/2}$ for $^{85}$Rb. Note that the two atomic species have the same angular momentum states.  For the Rb atoms, we used a 479.768~nm coupling laser; for the Cs atoms, we used a 510.018~nm coupling laser. We applied an $E$-field using a horn antenna placed 318~mm from the vapor cell.  Fig.~\ref{CsRbEIT} shows a typical simultaneous EIT signal measurement obtained from both the $^{133}$Cs and $^{85}$Rb atoms for a SG power of $-11$~dBm and at 9.222~GHz. We see that the measurement splitting ($\Delta f_m$) is different for the two atomic species, which is a result of the two atoms having different dipole moments (this is discussed in detail below).

If the RF is detuned from the on-resonant RF transition, the measurement splitting $\Delta f_m$ (or $\Delta f_o$) increases from the on-resonant AT splitting by the following \cite{rfdetune, berman}
\begin{equation}
\Delta f_{\delta}=\sqrt{\left(\delta_{RF}\right)^2+\left(\Delta f_{o}\right)^2}\,\,\, ,
\label{peaksep}
\end{equation}
where $\delta_{RF}$ is the RF detuning ($\delta_{RF}=f_{RF,o}-f_{RF}$; $f_{RF,o}$ is the on-resonance RF transition and $f_{RF}$ is the frequency of the RF source) and $\Delta f_o$ is the separation of the two peaks with no RF detuning (i.e., the on-resonant AT splitting or when $\delta_{RF}=0$). In order for us to compare measurements for the two different atomic species, we need to correct for the situation where the two species can have slightly different RF transition frequencies.  Alternatively, we can assume that the RF source produces the same $E$-field at the vapor cell for the slightly difference frequencies (within the $\% f$) and then perform on-resonant measurements for each of the two different species. We have verified that the RF source produces constant output power for a given $\% f$, and that the losses in the cable feeding the antenna and the antenna parameters are constant for a given $\% f$. This ensures that the $E$-field at the vapor cell is constant for a given $\% f$.  Therefore, we perform measurements at two slightly different RF frequencies (one at the on-resonant frequency for $^{133}$Cs, and one at the on-resonant frequency for $^{85}$Rb).  With that said, we need to ensure that the two frequencies are indeed at the on-resonant transition for the two atoms. While the data in Table \ref{t1} for $f_{RF,o}$ were calculated from the best current available quantum defects, there remains the possibility of errors in these quantum defects and in turn errors in the calculation of $f_{RF,o}$. As discussed in \cite{rfdetune}, an alternative approach for determining $f_{RF,o}$ is to perform RF detuning experiments and fit the expression in eq. (\ref{peaksep}) to a set of measurements for $\Delta f_{\delta}$ over a range of $\delta_{RF}$. This RF detuning data allows us to determine the on-resonant RF transitions (i.e., $f_{RF,o}$) {to within $\pm$~0.25 MHz (determined by averaging several sets of data)}. This measured $f_{RF,o}$ allows us to make comparisons to calculated values of $f_{RF,o}$ as determined from quantum defects; in effect, assessing the values of the current available quantum defects.

\begin{figure}
\centering
%\scalebox{.3}{\includegraphics*{Cs-RF-detuning-9p2GHz=With-fRF-QDefects_eps.eps}}\\
%{\tiny{(a)}}\\
%\scalebox{.3}{\includegraphics*{Rb-RF-detuning-9p2GHz=With-fRF-QDefects_eps.eps}}\\
%{\tiny{(b)}}\\
\scalebox{.3}{\includegraphics*{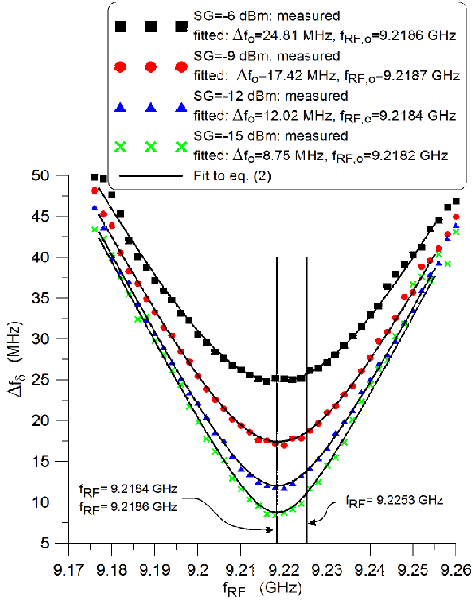}}\\
{\tiny{(a)}}\\
\scalebox{.3}{\includegraphics*{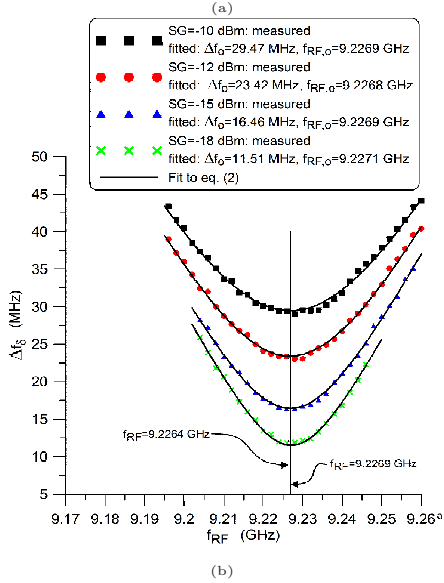}}\\
{\tiny{(b)}}\\
\caption{RF detuning experiments: (a) $^{133}$Cs: $43D_{5/2}$-$44P_{3/2}$  and  (b) $^{85}$Rb: $61D_{5/2}$-$62P_{3/2}$. The vertical lines and corresponding frequencies are the $f_{RF,o}$ obtained from both the RF detuning measurements and from the quantum defect data.}
\label{rfdetung1}
\end{figure}

As shown in eq. (\ref{peaksep}), a measurement for $\Delta f_o$ obtained from the off-resonant RF transition frequency will result in an over-estimate of $\Delta f_o$ and in turn an over estimate $|E|$. Thus, it is important that we determine the on-resonant transition frequencies. To determine these, we performed RF detuning experiments for various RF power levels ($P_{SG}$) for the two atomic species.  The data for $\Delta f_{\delta}$ for the two atoms are shown in Fig.~\ref{rfdetung1}. The data for both atoms were collected simultaneously.  Each curve for each atom was fitted to the expression in eq. (\ref{peaksep}) and the fitted $f_{RF,o}$ are shown in the figure. Averaging the data for the six different SG power levels, we find that $f_{RF,o}=9.2184$~GHz for $^{133}$Cs and $f_{RF,o}=9.2269$~GHz for $^{85}$Rb.  In order to compare these values to those given in Table \ref{t1} (i.e., the ones obtained from the quantum defects) we show vertical lines on Fig.~\ref{rfdetung1} indicating $f_{RF,o}$. For this set of Cs and Rb states we see that $f_{RF,o}$ obtained from the RF detuning experiments compare very well to those obtained from the calculations using the quantum defects, where the $^{133}$Cs results are in better comparison (every so slightly) than the $^{85}$Rb results.  In the next subsection we investigate a set of states where we see that there is a significant difference between $f_{RF,o}$ obtained from the calculated and detuning results.

On a side note, there are three papers that give quantum defects for Cs \cite{qdcs1, qdcs2, qdcsl}, where \cite{qdcs2, qdcsl} are sequential improvements over those in \cite{qdcs1}. If we used the quantum defect data given in \cite{qdcs1}, we obtain $f_{RF,o}$=9.2253~GHz (this frequency is shown in Fig.~\ref{rfdetung1}) which is not as close to the measured $f_{RF,o}$ and to those obtained from the new quantum defects \cite{qdcsl}. These RF detuning experiments help indicate the accuracy of the the very recent quantum defects given in \cite{qdcsl}, as well as those for Rb.

\begin{figure}
\centering
%\scalebox{.31}{\includegraphics*{Delta-f-Power-ScansCsandRb_eps.eps}}\\
\scalebox{.35}{\includegraphics*{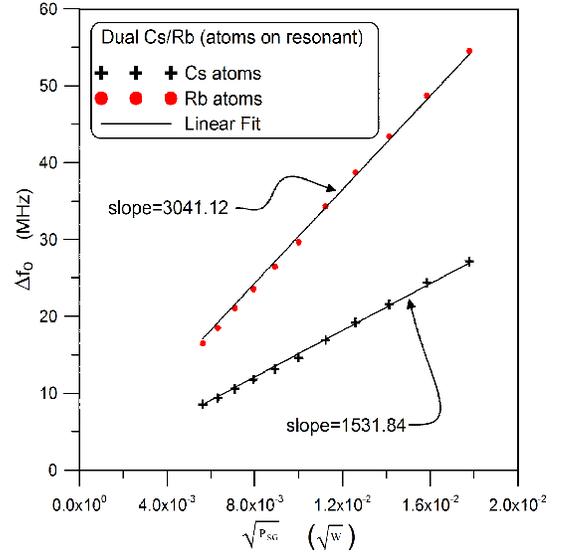}}\\
\caption{$\Delta f_o$ as a function of SG power for the $^{133}$Cs at $f_{RF}$=9.218~GHz  and  $^{85}$Rb at $f_{RF}$=9.227~GHz.}
\label{pscan9}
\end{figure}

With the on-resonant RF transition frequencies now determined, we then perform two sets of measurements for a range of SG power levels (one set for the on-resonant frequency of  $^{133}$Cs, or $f_{RF,o}=9.2184$~GHz; one set for the on-resonant frequency of  $^{85}$Rb, or $f_{RF,o}=9.2269$~GHz).   For the different SG powers we determine $\Delta f_o$ for the $^{133}$Cs for $f_{RF,o}=9.2184$~GHz and $\Delta f_o$ for the $^{85}$Rb for $f_{RF,o}=9.2269$~GHz. These results are shown in Fig.~\ref{pscan9}. All of these data were collected with all four laser beams propagating through the cell. That is, both atomic species where excited to high Rydberg states. Thus is further discussed below. From the figure, we notice that the slopes of each curve are different, which are determined by linear fit of the data and are shown in the figure.  This is expected because from eq. (\ref{mage2}), the measured splitting for each atom is proportional to $\wp\,|E|$. As discussed in \cite{r1}, this slope can be thought of as the measurement $E$-field sensitivity for a given atom.  Since each atom has a different dipole moment for their respective atomic states, we use (\ref{mage2}) to show that the ratio of the slopes for $\Delta f_o$ (for the same $E$-field seen by the two atoms) is given by
\begin{equation}
R=\frac{\wp_{Cs}}{\wp_{Rb}}=\frac{{\cal R}_{Cs}{\cal A}_{Cs}}{{\cal R}_{Rb}{\cal A}_{Rb}} \,\,\, ,
\label{RR}
\end{equation}
where $R$ is defined as the ``sensitivity ratio'' of $^{133}Cs$ to $^{85}Rb$; $\wp_{Cs}$ and $\wp_{Rb}$ are the dipole moments for $^{133}Cs$ to $^{85}Rb$, respectively. The assumption is that the same $E$-field at the vapor field is generated by the SG source for the two different closely-spaced frequencies (in this case 9.2184~GHz and 9.2269~GHz). This was verified by measuring both the output SG power and the loss in the cable. Using eq. (\ref{RR}) and the dipole moments given in Table \ref{t1}, we calculated the sensitivity ratio $R$ to be 0.505.  Using the data in Fig.~\ref{pscan9}, we determined the ratio of the slopes from the measurements (1531.84 for $^{133}$Cs and 3041.12 for $^{85}$Rb) to be 0.504.  The difference between the measured and theoretical values of the sensitivity ratio R is $0.1~\%$.  This helps confirm that the calculations of the two dipole moments for the two different atomic species are correct.
While a detailed uncertainties analysis for these type of measurements (including determining $\Delta f_o$, the slope of $\Delta f_o$, and the $E$-field strength) are currently being investigated, we have estimated that we can determined the slope of $\Delta f_o$ to within $\pm 0.4~\%$ (determined by averaging several sets of data).

\begin{figure}
\centering
%\scalebox{.31}{\includegraphics*{Power-ScansCsandRb-both-ON_eps.eps}}\\
\scalebox{.35}{\includegraphics*{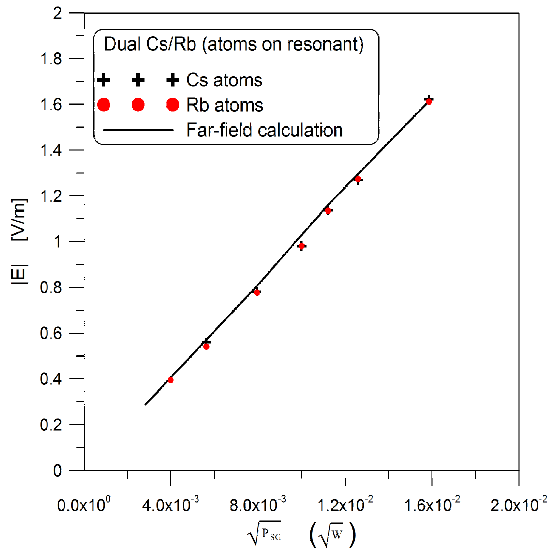}}\\
\caption{Calculated $|E|$-field as a function of SG power for the $^{133}$Cs at $f_{RF}$=9.218~GHz  and  $^{85}$Rb at $f_{RF}$=9.227~GHz.}
\label{efield9Ghz}
\end{figure}

The measured $\Delta f_o$ for each atomic species was used in eq. (\ref{mage2}) to calculate the $E$-field at the vapor cell for a range of $P_{SG}$. These calculated value are given in Fig.~\ref{efield9Ghz}. The $E$-field strength obtained from both the $^{133}$Cs and $^{85}$Rb atoms are the same. For a comparison, we estimated the E-field strength from a far-field calculation. Using $P_{SG}$, the cable loss (measured to be 2 dB), gain of the horn antenna (estimated to be 14.5 dBi), and the distance of the horn to the vapor ($x=31.8$ cm), we calculated the E-field in the far-field by \cite{stutzman}
\begin{equation}
|E|=\frac{\sqrt{59.96}}{x}\sqrt{10^{\frac{14.5}{10}}\,\,10^{0.001*\frac{(P_{SG}-2)}{10}}}\,\,\, ,
\label{friis}
\end{equation}
where $P_{SG}$ is given in units of  dBm. These far-field values are also shown in Fig.~\ref{efield9Ghz}.  The estimated E-field strength obtained for both Cs and Rb compare well to the far-field estimates.  This illustrates that the two different atomic species can be used simultaneously to independent measure the same E-field strength, resulting in two independent measurements of the $E$-field.

\begin{figure}
\centering
%\scalebox{.31}{\includegraphics*{Delta-f-Power-ScansCsandRb-ALL_eps.eps}}\\
\scalebox{.35}{\includegraphics*{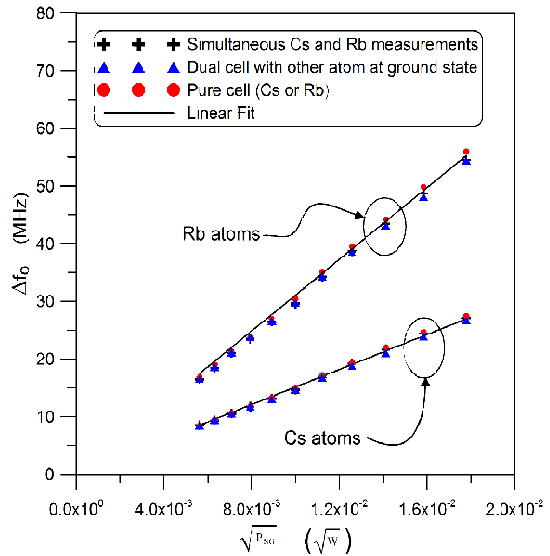}}\\
\caption{Comparison of measured $\Delta f_o$ with dual cell (with both atoms at Rydberg state or for one atom at ground state) and for a pure cell: $^{133}$Cs at $f_{RF}$=9.218~GHz  and  $^{85}$Rb at $f_{RF}$=9.227~GHz.}
\label{scancomp}
\end{figure}

{The presence of a second atomic system could affect the EIT measurements of the first atomic vapor. The two atomic species could interact with each other when they are both excited to high Rydberg state, or one species could possible act as a buffer gas from the other species’ perspective \cite{Sargsyan}. While others have observed interactions between Rb and Cs atoms \cite{krause}, these are at much higher vapor pressures than in our experiments, and as such, we do not expect to observe any effect on our measurements.}  To address these possibilities, we performed control experiments.  We first repeated the power scan for each atomic species separately.  That is, we performed measurements for $^{133}$Cs, while the probe and coupling lasers for $^{85}$RB were blocked from entering the vapor cell ($^{85}$Rb at ground state and $^{133}$Cs at a high Rydberg state). Similarly, we performed measurements for $^{85}$Rb, while the probe and coupling lasers for $^{133}$Cs were blocked from entering the vapor cell ($^{133}$Cs at ground state and $^{85}$Rb at a high Rydberg state). Secondly, we performed measurements with a pure $^{133}$Cs cell and with a pure $^{85}$Rb cell. These two cells were the same size as the two species cell. The data from all these different approaches (along with the data from above) are shown in Fig.~\ref{scancomp}.  The data show that all the approaches give the same value of $\Delta f_o$ and indicate that there is no significant interaction between the two different atomic species in the same vapor cell excited to high Rydberg states.

\subsection{$^{133}$Cs: $66S_{1/2}$-$66P_{3/2}$  and  $^{85}$Rb: $65S_{1/2}$-$65P_{3/2}$}

We next performed experiments for an RF transition at approximately 13.4~GHz. From Table \ref{t1}, that corresponds to $6S_{1/2}$-$6P_{3/2}$-$66S_{1/2}$-$66P_{3/2}$ for $^{133}$Cs and $5S_{1/2}$-$5P_{3/2}$-$65S_{1/2}$-$65P_{3/2}$ for $^{85}$Rb. Note that in this case the two atomic species have the same angular momentum states, but different angular states for the RF transitions than the previous case.  For the Rb atoms, we used a 479.718~nm coupling laser; for the Cs atoms, we used a 509.022~nm coupling laser. We applied a $E$-field via a horn antenna placed 415~mm from the vapor cell.

\begin{figure}
\centering
%\scalebox{.38}{\includegraphics*{Both-Atoms-RF-detuning-13p4GHz-fRF-Qdefects_eps.eps}}\\
\scalebox{.38}{\includegraphics*{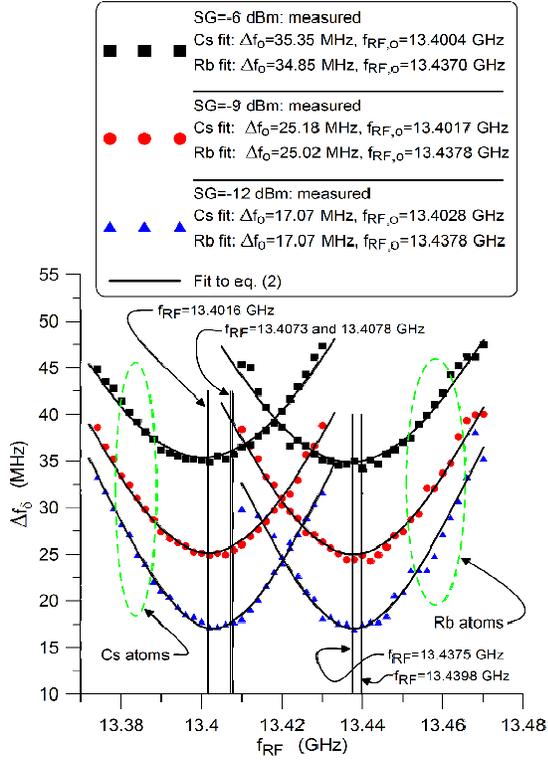}}\\
\caption{RF detuning experiments for both  $^{133}$Cs ($66S_{1/2}$-$66P_{3/2}$)  and  $^{85}$Rb ($65S_{1/2}$-$65P_{3/2}$). The vertical lines and corresponding frequencies are the $f_{RF,o}$ obtained from both the RF detuning measurements and from the quantum defects.}
\label{rfdetung2}
\end{figure}

Once again, to determine the on-resonant RF transition frequencies, we performed RF detuning measurements.  The data from these measurements are shown in Fig.~\ref{rfdetung2}. From a fitting of eq. (\ref{peaksep}) to these measurements and averaging the data for the different $P_{SG}$, we find that $f_{RF,o}=13.4016$~GHz for $^{133}$Cs and $f_{RF,o}=13.4375$~GHz for $^{85}$Rb.  These values along with the ones given in Table \ref{t1} (calculated from quantum defect data) are shown by vertical lines in Fig.~\ref{rfdetung2}. From this figure, we see that the calculated values for $f_{RF,o}$ are different by an appreciable amount. As such, if these calculated values for $f_{RF,o}$ are used for these measurements, then the $\Delta f_o$ obtained for the off-resonant RF transitions frequency will result in an over-estimate of $\Delta f_o$, and in turn an over estimate of $|E|$.

\begin{figure}
\centering
%\scalebox{.31}{\includegraphics*{Delta-f-Power-ScansCsandRb-13p4GHz_eps.eps}}\\
\scalebox{.35}{\includegraphics*{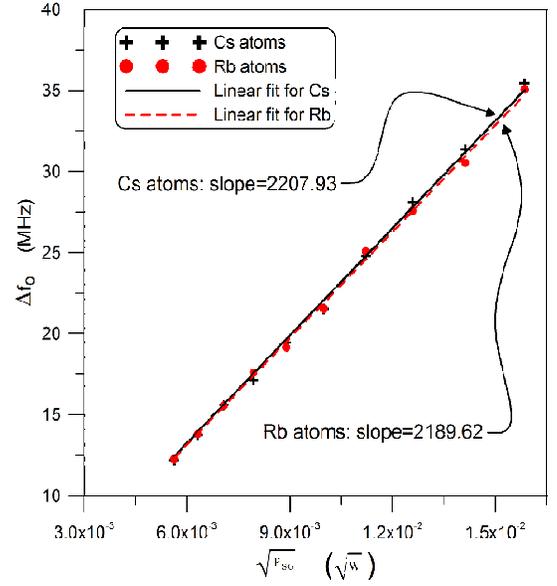}}\\
\caption{$\Delta f_o$ as a function of SG power for the $^{133}$Cs at $f_{RF}$=13.4016~GHz  and  $^{85}$Rb at $f_{RF}$=13.4375~GHz.}
\label{pscan13}
\end{figure}

With the on-resonant RF transition frequencies determined, we then performed two set of measurements for a range of $P_{SG}$ (one set for the on-resonant frequency of  $^{133}$Cs, or $f_{RF,o}=13.4016$~GHz; one set for the on-resonant frequency of  $^{85}$Rb, or $f_{RF,o}=13.4375$~GHz).  For the different SG powers, we determine $\Delta f_o$ for $^{133}$Cs at $f_{RF,o}=13.4016$~GHz and $\Delta f_o$ for $^{85}$Rb at $f_{RF,o}=13.4375$~GHz. These results are shown in Fig.~\ref{pscan13}. All the data were collected with all four laser beams propagating through the cell. That is, both atomic species were excited to high Rydberg states. Using the data in Fig.~\ref{pscan13}, we determined the ratio of the slopes from the measurements (2207.93 for $^{133}$Cs and 2189.62 for $^{85}$Rb) to be 1.008, and from eq. (\ref{RR}), the theoretical value for the sensitivity ratio $R$ (using the data in Table \ref{t1}) is 1.002.  The difference between the measured and theoretical values of the sensitivity ratio R is $0.6~\%$. This, once again, helps confirm that the calculations of the two dipole moments for the two different atomic species are correct.

\begin{figure}
\centering
%\scalebox{.31}{\includegraphics*{Efield-Power-ScansCsandRb_13p4GHz-All_eps.eps}}\\
\scalebox{.35}{\includegraphics*{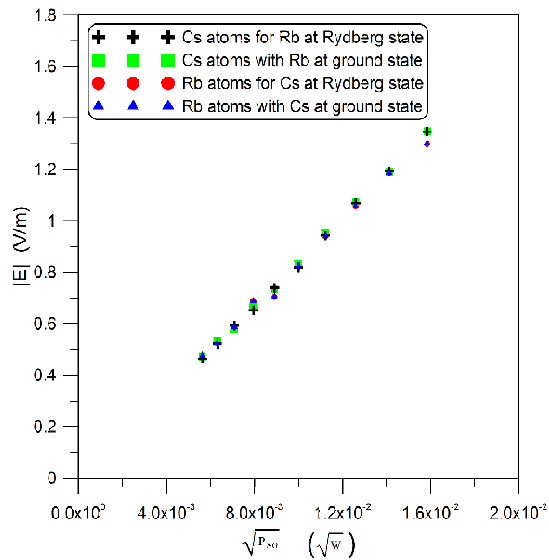}}\\
\caption{Calculated $|E|$-field as a function of SG power for the $^{133}$Cs at $f_{RF}$=13.4016~GHz  and  $^{85}$Rb at $f_{RF}$=13.4375~GHz. Comparison of measured $|E|$-field with  both atoms at Rydberg states and for one atom at ground state where no significant systematic atom-atom effects are observed.}
\label{efield13Ghz}
\end{figure}

The measured $\Delta f_o$ for each atomic species was used in eq. (\ref{mage2}) to calculate the $E$-field at the vapor cell for a range of SG power levels. These calculated values are given in Fig.~\ref{efield13Ghz}. The estimated E-field strength obtained for both Cs and Rb are the same.  To indicate that there is no interaction between each of these two highly excited Rydberg atoms, we repeated the power scan for each atomic species separately.  That is, we performed measurements for $^{133}$Cs, while the probe and coupling lasers for $^{85}$RB were blocked from entering the vapor cell (in effect, $^{85}$Rb at ground state and $^{133}$Cs at a high Rydberg state). Similarly, we performed measurements for $^{85}$Rb, while the probe and coupling lasers for $^{133}$Cs were blocked from entering the vapor cell (in effect, $^{133}$Cs at ground state and $^{85}$Rb at a high Rydberg state). Using these measured $\Delta f_o$, we calculated the $E$-field from eq.~(\ref{mage2}) and the results for these single atom measurements are also shown in Fig.~\ref{efield13Ghz}. The data shows that all the approaches give the same value of $|E|$-field and indicate that there are no significant interactions between two different atomic species in the same vapor cell excited to high Rydberg states.

\subsection{$^{133}$Cs: $40D_{5/2}$-$41P_{3/2}$  and  $^{85}$Rb: $68S_{1/2}$-$68P_{3/2}$}

Finally, we performed experiments for an RF transition of approximately 11.6~GHz. From Table \ref{t1} that corresponds to $6S_{1/2}$-$6P_{3/2}$-$40D_{5/2}$-$41P_{3/2}$ for $^{133}$Cs and $5S_{1/2}$-$5P_{3/2}$-$68S_{1/2}$-$68P_{3/2}$ for $^{85}$Rb. Note that unlike the two previous cases, in this case, the two atomic species have different angular momentum states for the RF transitions. For the Rb atoms, we used a 479.660 nm coupling laser; for the Cs atoms, we used a 510.302 nm coupling laser. We applied an $E$-field via a horn antenna placed 415~mm from the vapor cell.

\begin{figure}
\centering
%\scalebox{.30}{\includegraphics*{Delta-f-Power-ScansCsandRb-ALL-11p6GHz_eps.eps}}\\
\scalebox{.35}{\includegraphics*{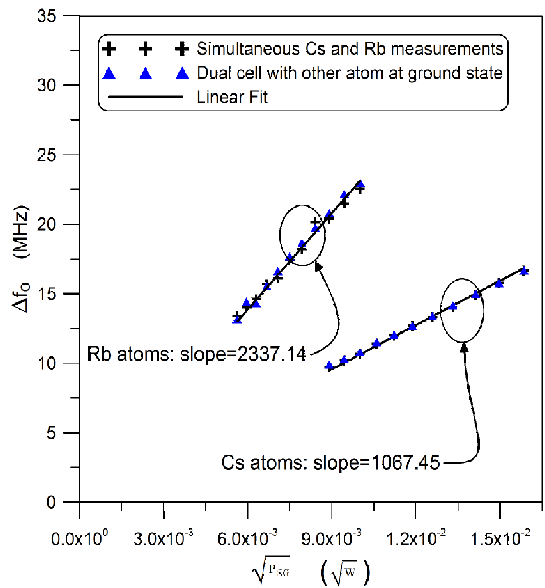}}\\
\caption{$\Delta f_o$ as a function of SG power for the $^{133}$Cs at $f_{RF}$=11.6172~GHz  and  $^{85}$Rb at $f_{RF}$=11.6656~GHz.}
\label{pscan11}
\end{figure}

Once again, to determine the on-resonant RF transition frequencies, we performed RF detuning measurements.  The data are not shown here, but we found that $f_{RF,o}=11.6172$~GHz for $^{133}$Cs and $f_{RF,o}=11.6656$~GHz for $^{85}$Rb.  With the on-resonant RF transition frequencies determined, we then performed two set measurements for a range of $P_{SG}$ (one set for the on-resonant frequency of  $^{133}$Cs, or $f_{RF,o}=11.6172$~GHz; one set for the on-resonant frequency of  $^{85}$Rb, or $f_{RF,o}=11.6656$~GHz).  These results for the measured $\Delta f_o$ for various power levels are shown in Fig.~\ref{pscan11}. All the data were collected with all four laser beams propagating through the cell. That is, both atomic species where excited to high Rydberg states. Using the data in Fig.~\ref{pscan11}, we determined the ratio of the slopes from the measurements (1067.45 for $^{133}$Cs and 2337.14 for $^{85}$Rb) to be 0.457, and from eq. (\ref{RR}), the theoretical value of sensitivity ratio $R$ (using the data in Table \ref{t1}) is determined to be 0.455.  The difference between the measured and theoretical values of the sensitivity ratio R is $0.4~\%$. This, once again, helps confirm that the calculations of the two dipole moments for the two different atomic species (each have different angular momentum in this case) are correct.

\begin{figure}
\centering
%\scalebox{.31}{\includegraphics*{Efield-Power-ScansCsandRb_11p6GHz-All_eps.eps}}\\
\scalebox{.35}{\includegraphics*{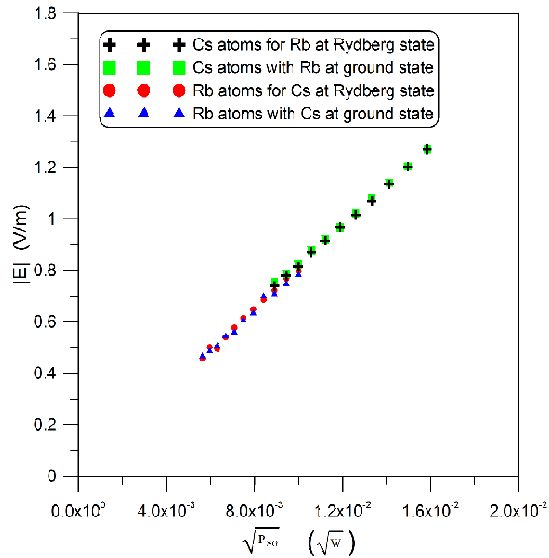}}\\
\caption{Calculated $|E|$-field as a function of SG power for the $^{133}$Cs at $f_{RF}$=11.6172~GHz  and  $^{85}$Rb at $f_{RF}$=11.6656~GHz. Comparison of measured $|E|$-field with  both atoms at Rydberg states and for one atom at ground state where no significant systematic atom-atom effects are observed.}
\label{efield11Ghz}
\end{figure}

To indicate that there is no significant interaction between each of these two highly-excited Rydberg atoms, we repeated the power scan for each atomic species separately.  That is, we performed measurements for $^{133}$Cs, while the probe and coupling lasers for $^{85}$RB were blocked from entering the vapor cell ($^{85}$Rb at ground state and $^{133}$Cs at a high-Rydberg state). Similarly, we performed measurements for $^{85}$Rb, while the probe and coupling lasers for $^{133}$Cs were blocked from entering the vapor cell ($^{133}$Cs at ground state and $^{85}$Rb at a high-Rydberg state). These measured $\Delta f_o$ are also shown in Fig.~\ref{efield11Ghz}. The data show that all the approaches give the same value for $\Delta f_o$ and indicate that there is no interaction between the two different atomic species in the same vapor cell excited to high Rydberg states.

The measured $\Delta f_o$ for each atomic species were used in eq. (\ref{mage2}) to calculate the $E$-field at the vapor cell for a range of $P_{SG}$. These calculated values are given in Fig.~\ref{efield11Ghz}. The estimated E-field strength obtained for both Cs and Rb compare well.  This illustrates that the two different atomic species can be used simultaneously to independently measure the same E-field strength, resulting in two independent measurements of the $E$-field.  Also shown in this figure are the results for the two different atoms at ground states, indicating no significant Rydberg atom interactions.

These results for the E-field illustrate the interesting point that using two atomic species simultaneously, one can expand the range of the measurements. Rb atoms have difficulty measuring $E$-fields for high $P_{SG}$, and the Cs atoms have difficultly measuring $E$-fields for lower $P_{SG}$.  For low E-fields strength it is difficult to measure and/or detect splitting in the EIT signal. Since the measured $\Delta f_m$ (or $\Delta f_o$) is directly proportional to the product of ``$\wp\,|E|$'' (i.e., eq. (\ref{mage2}), when the E-field strength is weak and $\wp$ is small, the ability to measure $\Delta f_m$ becomes problematic).  For this particular set of $^{133}$Cs and $^{85}$Rb states, the dipole moment for $^{85}$Rb is twice as large as the dipole moment for $^{133}$Cs, and hence the $^{85}$Rb atoms can measure a $50~\%$ weaker field. This is evident in Fig.~\ref{efield11Ghz} where the smallest field for the $^{133}$Cs atoms that can be detected is 0.8~V/m and the smallest field for the $^{85}$Rb atoms that can be detected is 0.4~V/m.  The maximum detectable field for each atom is limited partially by the methods in which the probe lasers are scanned. We use acoustic-optic modulators (AOMs) to scan the probe laser and as such once AT splitting (i.e., $\Delta f_m$) becomes greater than the AOM scan range, an $E$-field cannot be detected.  Since the dipole moment for $^{85}$Rb is twice as large as the dipole for $^{133}$Cs for this particular set of $^{133}$Cs and $^{85}$Rb states, the $^{85}$Rb atoms will reach this scan limit first, as indicated in the figure.

\section{Conclusions}

In this paper, we demonstrated simultaneous $E$-field measurement via EIT using both cesium and rubidium in the same vapor cell.
Performing such a dual experiment helps quantify various aspects of this type of $E$-field metrology approach, which are important to understand when establishing an international measurement standard for an E-field strength and is a necessary step for this method to be accepted as a standard calibration technique.  For example, these experiments help in assessing the accuracy in the calculation of the dipole moment of the various atoms, where we showed the difference between the measured and theoretical values of the sensitivity ratio R was $0.6~\%$ or less for the three cases given here.  This dual atomic species experiment also allows us to investigate the possibility that the two atomic species could interact with each other when they are both excited to high-Rydberg states. To address this possibility, we performed a set of experiments in a pure vapor cell, and two separate experiments in a cell with two atomic species. In the separate dual cell experiments, we performed measurements on one atomic species with the probe and coupling lasers for other atomic species blocked from entering the vapor cell (in effect, one atom at ground state and the other at a high-Rydberg state).  From these experiments, the two different atomic species appear to not significantly interact when they are both excited to high-Rydberg states (at least for the types of measurements of interest in this paper).  Finally, the RF detuning results presented here also help quantify the accuracy of reported quantum defects, which are used in various aspects of these types of measurements.

\section{Acknowledgements}
We thank Dr. Georg Raithel and Dr. David A. Anderson of the University of Michigan for their useful technique discussions. This work was partially supported by the Defense Advanced Research Projects Agency (DARPA) under the QuASAR Program and by NIST through the Embedded Standards program.

\end{document}